\documentclass[aps,pre,twocolumn,showpacs,superscriptaddress,groupedaddress]{revtex4-2}
\usepackage{graphicx}  % needed for figures
\usepackage{dcolumn}   % needed for some tables
\usepackage{bm}        % for math
\usepackage{dsfont}    % for math (double-stroked characters via \mathds{text})
\usepackage{amssymb}   % for math
\usepackage{amsmath}
\usepackage{journals}
\usepackage{color}
\usepackage{siunitx}
\usepackage{physics}
\usepackage{hyperref}

\usepackage{accents}
\newcommand\undervec[1]{\underaccent{\vec}{#1}}
\DeclareMathOperator{\Li}{Li}

\begin{document}

\title{Topological Design of Heterogeneous Self-Assembly}
\author{Andrei A. Klishin}
\email{aklishin@seas.upenn.edu}
\altaffiliation[Current address:]{Department of Bioengineering, School of 
Engineering \& Applied Science, University of Pennsylvania, Philadelphia, PA 
19104, USA}
\affiliation{Department of Physics, University of Michigan, Ann Arbor, Michigan 
48109, USA}
\affiliation{Center for the Study of Complex Systems, University of Michigan, 
Ann Arbor, Michigan 48109, USA}
\affiliation{John A. Paulson School of Engineering and Applied Sciences, 
Harvard University, Cambridge, MA 02138, USA}

\author{Michael P. Brenner}
\affiliation{John A. Paulson School of Engineering and Applied Sciences, 
Harvard University, Cambridge, MA 02138, USA}

\date{\today}

\begin{abstract}
Controlling the topology of structures self-assembled from a set of 
heterogeneous building blocks is highly desirable for many applications, but is 
poorly understood theoretically. Here we show that the thermodynamic theory of 
self-assembly involves an inevitable divergence in 
chemical potential.  The divergence 
 and its detailed structure are controlled by the spectrum of the 
transfer matrix, which summarizes all of self-assembly design degrees of 
freedom. By analyzing the transfer matrix, we map out the phase boundary 
between the designable structures and the unstructured aggregates, 
driven by the level of cross-talk.
\end{abstract}

\maketitle
%\section{Introduction}
%self-assembly is powerful
Self-assembly is an robust method for organizing 
loose building blocks to adopt desired structures\cite{whitesides02}. A wide 
range of structures has been observed in synthetic self-assembly in both 
experiments and numerical simulations, from strictly periodic crystals 
\cite{zoopaper} to quasicrystals \cite{amirnature}, liquid 
crystals \cite{mohraz2005direct}, polymers \cite{ilievski2011self}, 
nets \cite{niu2019magnetic}, and finite clusters \cite{meng2010clusters}. While 
initial investigations were guided by anecdotal evidence and limited 
availability of building blocks, later works deliberately optimize 
 building block properties to assemble a desired target 
structure \cite{digitalalchemy}. The emergent connection between  building 
block properties and the structures they form makes self-assembly a 
model system for discussion of more general design 
principles \cite{sherman2020inverse}.

\begin{figure}
	\begin{center}
		\includegraphics[width=.5\textwidth]{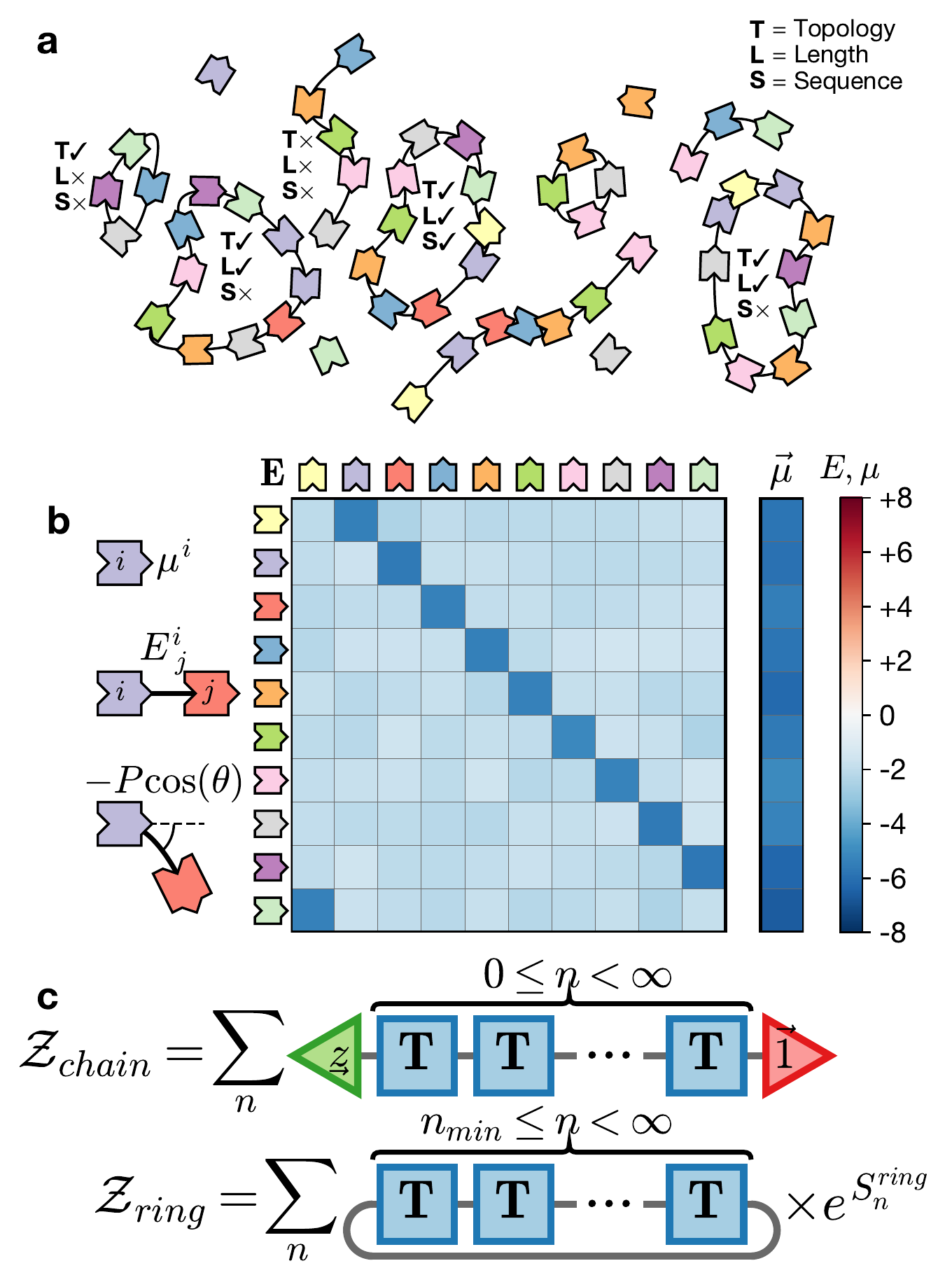}
	\end{center}
	\caption{The design space of the heterogeneous self-assembly problem. (a) 
		Self-assembled structures vary by topology (T), length (L), and 
		sequence 
		(S). Check and cross marks indicate whether the structures match the 
		target 
		(here, a length 10 ring). (b) The set of building blocks is 
		characterized 
		by their chemical potentials $\mu^i$, key-lock binding energies 
		$E^i_{\;j}$, and bending rigidity $P$. The on-target binding energies 
		are 
		very negative (dark blue), while the off-target binding energies are 
		only 
		slightly negative (light blue). (c) Computations of partition functions 
		for 
		linear chains and closed rings in graphical (tensor) notation, 
		corresponding to Eqns.~(\ref{eqn:Zchain}--\ref{eqn:Zring}).
	}
	\label{fig:space}
\end{figure}

%why we want topology
A great challenge in self-assembly design is controlling the topology of target 
structures (Fig.~\ref{fig:space}a). Topologically closed structures, such as 2D 
rings and 3D spheres \cite{dinsmore2002colloidosomes,sun2015ring}, have 
important applications in functional foods \cite{gibbs1999encapsulation}, 
biomedicine \cite{lanza2020principles}, and drug delivery 
\cite{rivas2017nanoprecipitation}, and it is natural to ask how this can be 
encoded into interactions between components. 
The competition between open and closed structures at finite concentration is naturally 
described by grand canonical thermodynamic theory. However,  such a theory has an inevitable divergence 
in the chemical potential. In this paper, we use this divergence as  a design tool to encode self assembled structure. We use this to make quantitative 
predictions of self-assembly yields of linear chains and closed rings for 
\emph{arbitrary} binding energy matrices, and then use this to derive  qualitative and 
quantitative design rules for \emph{optimal} binding energy matrices. This allows 
predicting the limits on complexity of structures driven by interaction 
cross-talk. This theoretical framework accounts for a whole set 
of building blocks, and thus opens the design of even more complex, 
higher-dimensional, branching, and loopy structures.

%\section{Design space and transfer matrix}

%\subsection{Design space}
In order to self-assemble target structures, we seek to jointly design a 
set of $N$ building blocks, where each building block has a ``lock'' and ``key''  binding site on 
 opposite sides. Binding occurs between the locks and keys of different particles. Lock and 
key interactions have been experimentally realized \cite{pacman,lockkey3d,lockkeykin} but for the present investigation we 
abstract  specific realizations. The set of 
building blocks is then described by the chemical potential 
concentration vector 
$\vec{\mu}$, the binding energy matrix $\mathbf{E}$, and the bond rigidity 
$P$ (see Fig.~\ref{fig:space}b).

Building block interaction is controlled by the 
binding energy. We denote the binding energy between the key side of particle 
$i$ and the lock side of particle $j$ as $E^i_{\;j}$, with  indices $i,j$ spanning
the range $[1,N]$; the entries form a matrix $\mathbf{E}$ (see 
Fig.~\ref{fig:space}b). To
design $\mathbf{E}$, we designate certain $(i,j)$ pairs as 
on-target, by making  the corresponding binding energies  
stronger (more negative) $\abs{E_\textsf{on-target}}\gg \abs{E_\textsf{off-target}}$.

Even with designed interactions, off target binding is ubiquous
in every practical experimental system, including DNA 
origami \cite{rothemund2006folding}, DNA-coated 
colloids \cite{macfarlane2011nanoparticle}, \emph{de novo} 
proteins \cite{huang2016coming}, and magnetic panels \cite{niu2019magnetic}. 
This means we cannot control all of the entries 
of $\mathbf{E}$ independently: adding an interaction patch with a 
specific on-target partner requires addding the off-target bindings as well.  
Off-target interactions limit the effective number of distinct building blocks 
in the system and thus the yield of complex self-assembled 
structures \cite{huntley2016information, murugan2015undesired}.  In this paper 
we 
study how  off-target binding specifically limits our ability to control 
topology via the binding energy matrix, with the bond rigidity and the relative 
concentrations of building blocks serving as secondary 
parameters \cite{murugan2015undesired}.

%\subsection{Transfer matrix combinatorics}
To proceed, we construct 
the \emph{transfer matrix} $\mathbf{T}$, which is  defined element-wise 
following Ref.~\cite{murugan2015undesired}:
\begin{equation}
	T^i_{\;j}=\exp(\beta(-E^i_{\;j}+\mu^j)+S_b).
	\label{eqn:T}
\end{equation}

The transfer matrix corresponds to adding one more element to the chain of 
building blocks. The first index $i$ accounts for the possible key binding 
sites exposed by the preceding part of the chain, while the second index $j$ accounts 
for binding to locks.
This definition of the transfer matrix closely follows that used in 
solving spin lattice models \cite{goldenfeld}, with the chemical potential 
playing the role of external field. It is also very similar to the 
coupling tensors in Refs.~\cite{spacenetwork,elo}. In the present problem, the topology of interactions is  simple, so the matrix product and 
tensor contraction views are  equivalent.  In what follows, we use this tensor perspective 
to compute the partition functions for all possible chains and rings (see 
Appendix~\ref{app:combin} for a full derivation). The expressions 
are presented 
in a graphical tensor form in Fig.~\ref{fig:space}c, but can also be written 
down analytically in closed form:
\begin{align}
	\mathcal{Z}_{chain}&=\undervec{z}\mathbf{D} \vec{1} \label{eqn:Zchain}\\
	\mathcal{Z}_{ring}&=\frac{1}{(2\pi\xi)^{3/2}}\left( 
	\Tr(\Li_{3/2}(\mathbf{T})) 
	-\sum\limits_{n=1}^{n_{min}-1}\frac{\Tr(\mathbf{T}^n)}{n^{3/2}}
	\right),\label{eqn:Zring}
\end{align}
where $\mathbf{D}\equiv\left(\mathbf{I}-\mathbf{T}\right)^{-1}$ is the 
\emph{propagator}, or sum of a matrix geometric series, $\Li_q(x)$ is the 
$q$-order polylogarithm function generalized to matrix arguments, and $\xi$ is 
the persistence length of the polymer chain, related to its rigidity $P$. The 
key difference between the two expressions is that self-assembly of closed 
rings requires paying a hefty \emph{entropic loop penalty} so that the end of 
the closed chain is co-located with the beginning \cite{doiedwards1988}. This 
penalty significantly 
reduces the assembly rate of the rings.

%\section{Divergence is inevitable}
\begin{figure}
	\begin{center}
		\includegraphics[width=.5\textwidth]{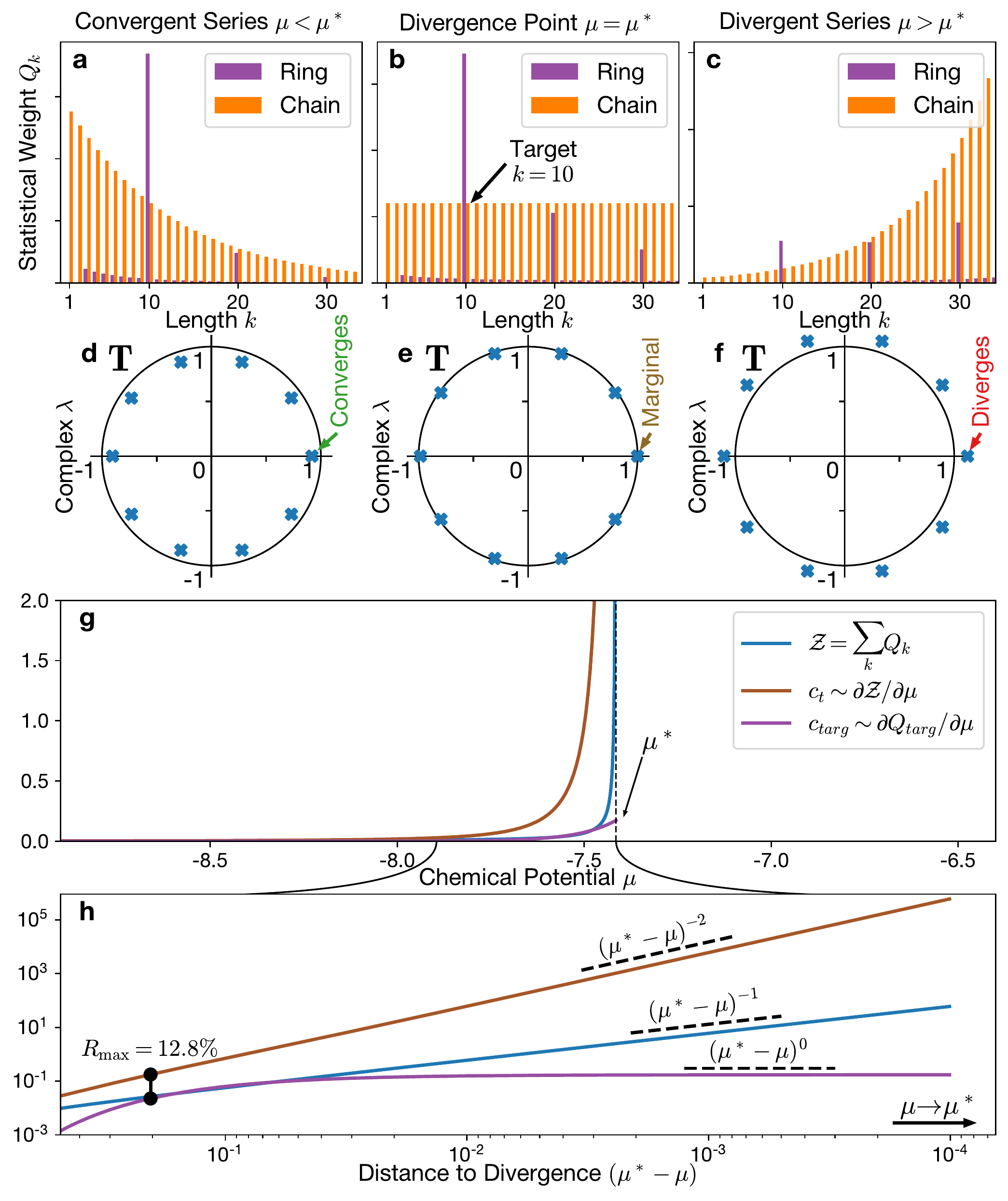}
	\end{center}
	\caption{Design of self-assembling clusters needs to hide structure in 		
	nearly-divergent series. (a,b,c) Sequences of statistical weights $Q_k$ in 
	a system designed well to assemble rings of length $k=10$, at different 
	$\mu$. (d,e,f) Complex 
	eigenvalue spectra of the corresponding transfer matrices $\mathbf{T}$, 
	with the top eigenvalue $\lambda_0$ determining the divergence. (g) Graph 
	of partition function $\mathcal{Z}$, total concentration $c_{t}$, and 
	target concentration $c_{targ}$ against the chemical potential $\mu$ in 
	\emph{linear} axis scales. The point labeled $\mu^*$ indicates the 		
	divergence. (h) Graph of the same quantities as (g) but in \emph{log-log} 	
	axis scales in vicinity of $\mu^*$. Dashed lines guide the eye for 
	different power law divergence scaling. Maximal conversion ratio 
	$R_{\max}$ 
	is achieved at finite distance from the divergence.}
	\label{fig:divergence}
\end{figure}

The expressions \eqref{eqn:Zchain}-\eqref{eqn:Zring} have a form specific to 
our building blocks. More generically, the combinatorial partition 
function of any sticky particle model takes the form of an infinite series of 
statistical weights $Q_s$ of all possible structures $s$:
\begin{equation}
	\mathcal{Z}=\sum\limits_{s}\exp(\beta\left(\undervec{n}_s \cdot \vec{\mu} - 
	E_s\right)+S_s),
	\label{eqn:Qs}
\end{equation}
where $\undervec{n}_s$ is the vector of numbers of particles of each type in 
the structure $s$, $E_s$ is the energy of all bonds, and $S_s$ is the internal 
entropy of the cluster due to vibrations and rotations. The sum over 
cluster types $s$ typically has an infinite number of terms with ever 
increasing particle counts $\undervec{n}_s$. Although there is a parameter regime where this infinite sum converges,  since the chemical potentials 
$\vec{\mu}$ are free parameters, we can always 
make them 
arbitrarily high, rendering the series 
divergent. While such a divergence is generic for all partition 
functions with an infinite number of terms, the exact value of ${\mu^i}^*$ at 
which the divergence occurs, as well as the behavior of the partition function close to the 
divergence is model-specific.

For the specific building block model discussed here, the typical 
sequences of statistical weights $Q_k$ in terms of structure length $k$ are 
illustrated in Fig.~\ref{fig:divergence}a-c. Depending on the chemical 
potential $\mu$, the series can either converge (panel a), diverge (panel c), 
or be exactly at the margin (panel b). Since the partition functions for chains 
and rings \eqref{eqn:Zchain}-\eqref{eqn:Zring} are computed as infinite series 
in matrices, their convergence is determined by the eigenvalue spectrum of the 
transfer matrix $\mathbf{T}$. While  the eigenvalues are generally complex, the top eigenvalue $\lambda_0$ is guaranteed to be real by the 
Perron-Frobenius theorem. The partition function converges when $\lambda_0<1$ 
(panel d of Fig.~\ref{fig:divergence}), diverges for $\lambda_0>1$ (panel f), 
and switches between the two regimes at $\lambda_0=1$ (panel e).

Is this divergence a physically observable phenomenon or a phase transition? 
The divergence occurs under variation of the chemical potential $\mu$, 
directly related to the concentration of \emph{free} (unbound) building blocks 
$c_{free}=c_0 e^{\beta \mu}$. In experiments we do not control the 
\emph{free} concentration, but the \emph{total} concentration $c_t$ of building 
blocks. 
In particular, we can take derivatives of the partition functions 
\eqref{eqn:Zchain}--\eqref{eqn:Zring} with respect to $\mu$ and compute 
similar, albeit bulkier, 
closed form expressions for the concentrations of building blocks bound in all 
structures $c_t$, or only in the desired target structure $c_{targ}$. We 
present the values of $\mathcal{Z}$,$c_t$,$c_{targ}$ in 
Fig.~\ref{fig:divergence}g-h, first in linear scale to emphasize the 
divergence, and then in log-log scale to emphasize the scaling.

In vicinity of the divergence, the lead eigenvalue scales as 
$\lambda_0\simeq e^{\beta(\mu-\mu^*)}$, dominating the partition function, and  determining the divergence exponents for both the 
partition functions and the total concentrations 
(Fig.~\ref{fig:divergence}g-h). We find that the total 
concentrations of building blocks bound in chains and in rings have the same 
divergence point $\mu^*$ but different asymptotic behaviors \cite{polylog}:
\begin{align}
	c_{chain}&\sim (\mu^*-\mu)^{-2}\\
	c_{ring}&\sim (\mu^*-\mu)^{-1/2}.
\end{align}

These expressions demonstrate that arbitrarily close to the divergence, i.e. at 
high concentrations, many more building blocks are bound chains than in rings, 
{\sl regardless of our design efforts}. Moreover, we find that near the divergence, the total 
concentrations are dominated by arbitrarily long chains and rings, longer than 
any finite target we might choose. Therefore, the highest conversion ratio of 
raw 
monomers into the designed finite-size targets  always occurs a finite distance 
from the divergence point. To quantify this in what follows, we use the ratio $R\equiv c_{targ}/c_t$ .

%\section{Mitigating the divergence}

\begin{figure}
	\begin{center}
		\includegraphics[width=.5\textwidth]{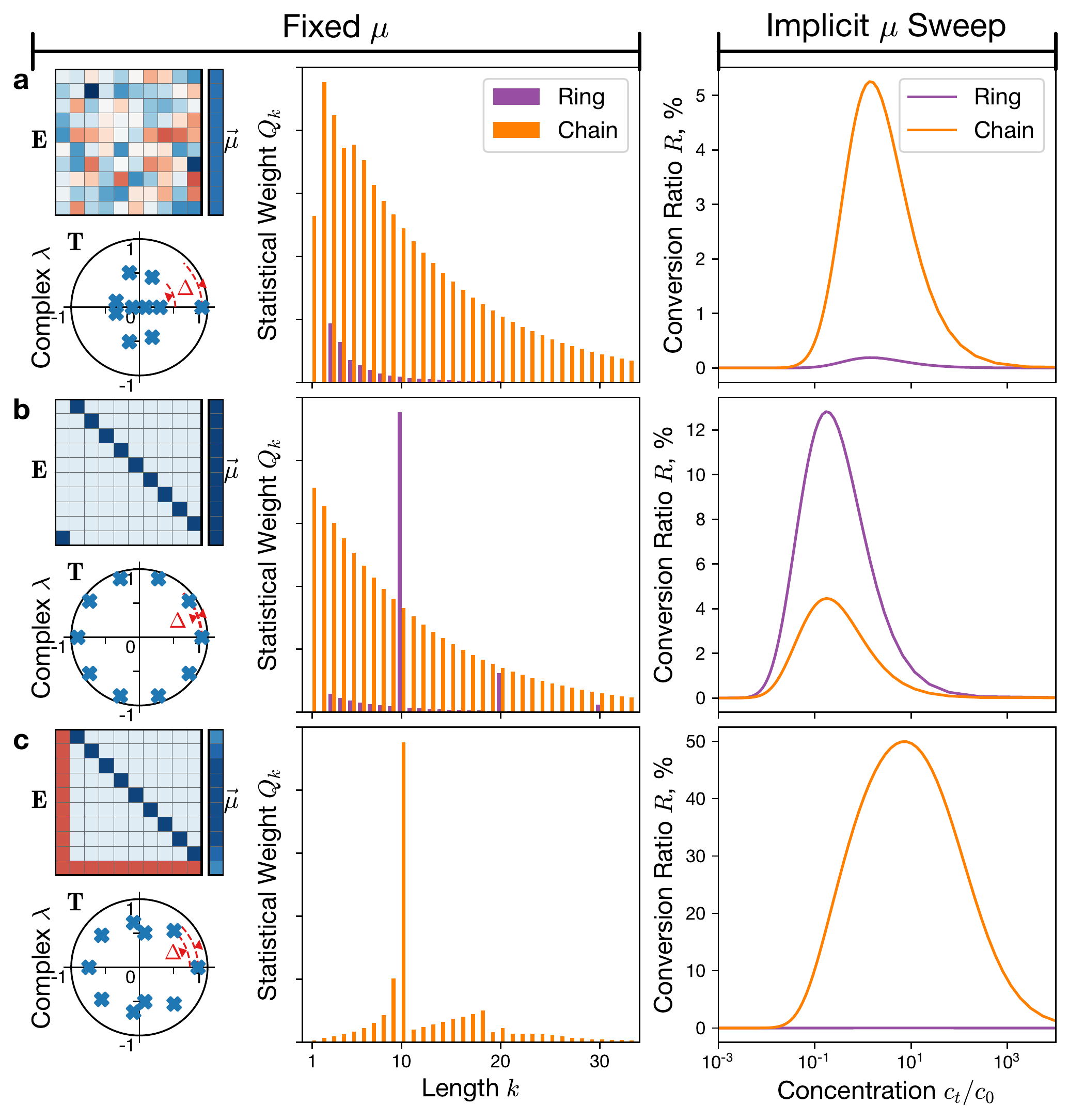}
	\end{center}
	\caption{Examples of yield predictions for several building block set 	
		designs. Columns of panels show different metrics, rows show different 
		designs. Columns: (left-top) Binding energy matrix $\mathbf{E}$ and 
		chemical potential vector $\vec{\mu}$. (left-bottom) Complex spectrum 
		of 
		the corresponding transfer matrix $\mathbf{T}$. Red arcs indicate the 
		spectral gap $\Delta$ between the first and second eigenvalues. 
		(center) 
		Statistical weight $Q_k$ sequences. (right) Conversion ratio of raw 
		monomers into the target structure, summed over all monomer types, 
		plotted 
		as a parametric curve with implicit parameter $\mu$. Rows: (a) Random 
		unstructured binding energy matrix. (b) Binding energy matrix designed 
		for 
		assembly of rings of length $k=10$. (c) Binding energy matrix and the 
		chemical potential profile designed for assembly of chains of length 
		$k=10$.}
	\label{fig:scenarios}
\end{figure}

Fig.~\ref{fig:scenarios} shows how to work around this divergence and 
make testable experimental predictions. We first design the binding energy 
matrix $\mathbf{E}$ and the chemical potential profile $\Delta \mu$ at fixed 
$\mu$ and study the eigenvalue spectrum of $\mathbf{T}$ and the sequence of 
$Q_k$ (left part of figure). We then sweep over $\mu$ as an implicit parameter, 
compute $c_t(\mu)$ and $R(\mu)$ and plot them against each other (right 
part of figure, see Appendix~\ref{app:combin} for details).

%overview - N and k = 10, absorb Sb, set xi=1
To unravel the connection between design 
space 
and self-assembly yields, we use the analytic expressions to explore several 
specific design scenarios. In all three cases we study a set of $N=10$ building 
block types and focus on the yields of structures of length $k=10$, both chains 
and rings. We choose the value of correlation length $\xi=1$. We keep 
$\beta=1$ and vary the binding energy 
matrix to explore three design scenarios 
shown in Fig.~\ref{fig:scenarios}.

%scenario unstructured, generic formation of aggregates, can be large but 
%unstructured
Fig.~\ref{fig:scenarios}a shows a random binding energy matrix $\mathbf{E}$. This occurs in practice
 when ``on-target'' and ``off-target'' bindings 
are effectively indistinguishable. The sequence of statistical weights $Q_k$ 
then closely resembles a simple unstructured geometric series. This does not 
mean that clusters do not form -- instead, the formed clusters can be large but 
don't have any particular structure. This is a \emph{failure state} of design, 
generically observed when off-target interactions are strong.

%scenario rings - small eigengap
Fig.~\ref{fig:scenarios}b shows a binding energy matrix $\mathbf{E}$ designed 
for forming closed rings. The key of each building block $i$ has strong 
on-target binding to the next lock $i+1$, with the last one binding to the first 
in a cyclical fashion. In the graphical representation of the matrix, this is 
visible as a dark first superdiagonal and an opposite corner element. The 
complex spectrum for this scenario shows a very small spectral gap $\Delta$, which 
as we show below is a strong predictor and a requirement of forming rings of 
precise sequence. In the sequence of statistical weights this design causes a 
strong peak for $k=10$ rings and smaller ones for $k=20$ and $k=30$ (when the 
full ring sequence is repeated multiple times). This design also shows a 
substantial $R\approx 12\%$ for the rings despite them being fundamentally 
disadvantaged by the loop entropy penalty.

%scenario chains - modulation of mu, note peak density is higher
Fig.~\ref{fig:scenarios}c shows a system of binding energy matrix $\mathbf{E}$ 
and chemical potential vector $\vec{\mu}$ designed for forming finite chains. 
In the binding energy matrix $\mathbf{E}$, the left column and the bottom row 
are red (positive, repulsive), corresponding to the removed lock binding site 
on the first building block and key binding site on the last one. In the 
chemical potential vector, the entries for the first and last building blocks 
are substantially larger than the ones in the middle of the vector following 
 Ref.~\cite{murugan2015undesired}. As a 
result, in the sequence of statistical weights $k=10$ is much larger than any 
of the other ones. Almost $R\approx 50\%$ of building blocks are converted into 
length 10 chains.

%this was intuitive guesses, need to clarify the limits and formulate the 
%design rules
The designs presented in Fig.~\ref{fig:scenarios} are intuitive guesses, picked 
for illustrative purposes. Any real system of heterogeneous building blocks 
is highly constrained by the off-target binding energy, limiting the size of 
structures that can be reliably self-assembled. For any self-assembly theory to 
be useful in practice, the design rules must be not just 
qualitative, but quantitative. The design rules for open chains have been 
studied previously in Ref.~\cite{murugan2015undesired}, so here we focus on the 
design rules for closed rings and derive the limits on structure size set by 
cross-talk. We consider cyclic binding matrices of form shown in 
Fig.~\ref{fig:scenarios}b with variable on-target $v$ and off-target $\epsilon$ 
binding energies such that $v<\epsilon<0$.

%\section{Design rules}
\begin{figure}
	\begin{center}
		\includegraphics[width=.5\textwidth]{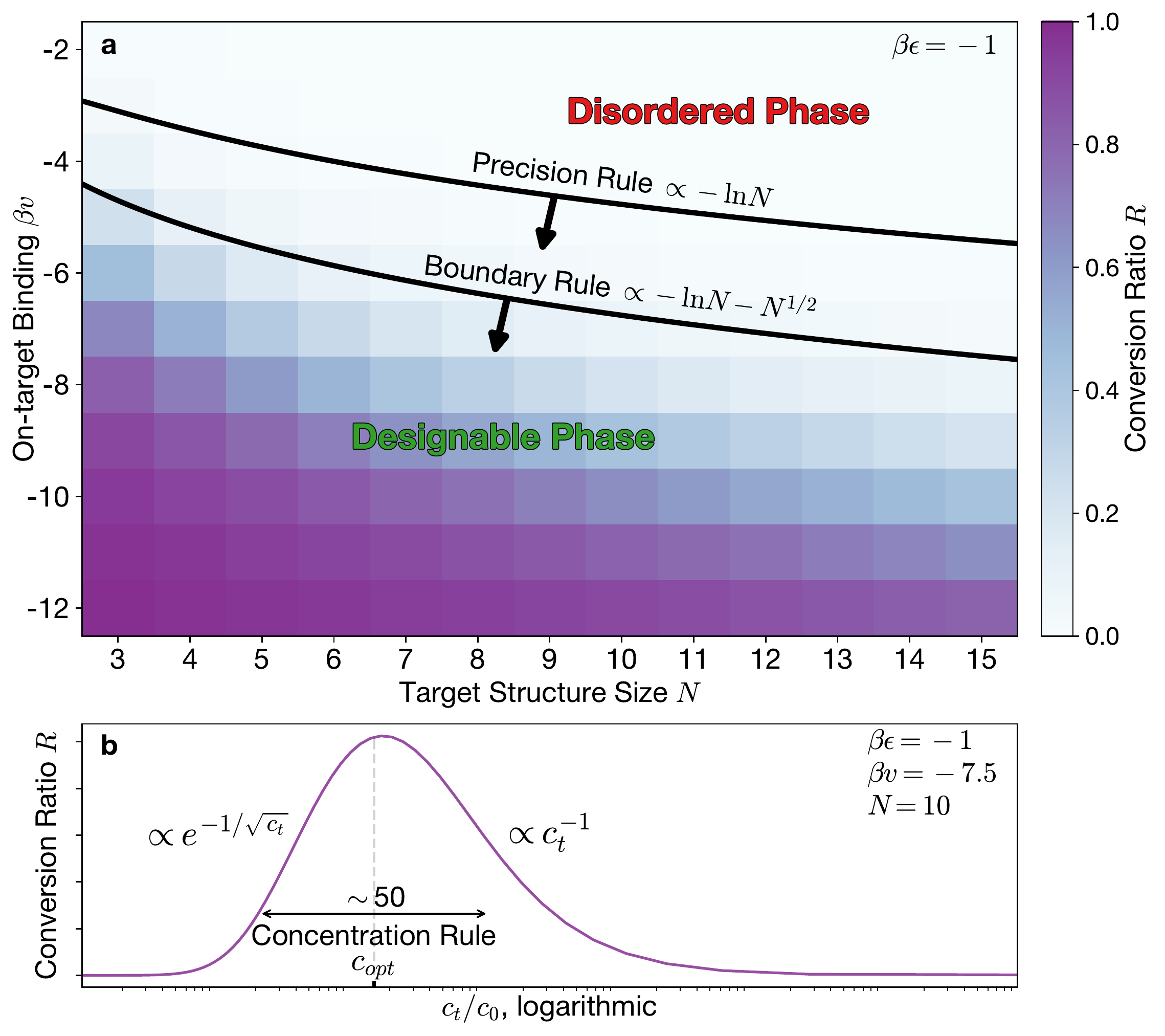}
	\end{center}
	\caption{Design rules for self-assembly of closed rings. (a) Heatmap of 
	conversion ratio $R$, maximized over the implicit chemical potential $\mu$. 
	Horizontal axis is the number of building block types, equal to the target 
	structure size. Vertical axis is the strength of on-target binding $\beta 
	v$ with off-target binding fixed at $\beta \epsilon=-1$. Black curves 
	illustrate the precision and boundary design rules, which separate the 
	amyloid and the designable phases. (b) 
	Typical curve of conversion ratio $R$ against the total concentration 
	$c_t$. The curve maximum lies around the estimated value $c_{opt}$, and the 
	logarithmic width of the high-yield regime is around $e^4\simeq 50$ fold. 
	The shapes of left and right curve tails have different asymptotic 
	behaviors, as indicated.}
	\label{fig:rules}
\end{figure}

Our results suggest a series of design rules:
First, the \emph{precision rule} ensures that the building 
blocks assemble the proper linear sequence. When this occurs, the weight $Q_k$ 
is significantly boosted for $k=N$ (see Appendix~\ref{app:desrules} for 
derivation). This boost is 
significant when the eigengap $\Delta=\abs{\lambda_0}-\abs{\lambda_1}$ is 
sufficiently small. This is achieved when on-target binding overcomes the many 
possible off-target  bindings:
\begin{equation}
	\beta(v-\epsilon)\ll -\ln(\frac{N^2}{\ln N}).
	\label{eqn:prec_ring}
\end{equation}

This design rule is qualitatively similar though  slightly more stringent than the ``$\ln N$ 
rule'' for linear chains derived in Ref.~\cite{murugan2015undesired}. If this 
constraint on the binding energy \emph{contrast} $\epsilon-v$ is not observed, 
the assembled chains will have random sequences 
(Fig.~\ref{fig:scenarios}a), and controlling any higher-level structure would 
become impossible.

The second design rule is the \emph{boundary rule}, ensuring that the last building block in the 
ring sequence binds to the first one. For this to occur, the binding energy 
needs to be strong enough to overcome the ring entropy penalty. Assuming the 
boundary rule \eqref{eqn:prec_ring} holds, we derive the following, stronger 
condition:
\begin{equation}
	\beta v\ll S_b - 
	\frac{3}{2}\ln(2\pi\xi)-\frac{3}{2}\ln(N)-\frac{e^{-\beta 
			\epsilon}}{(2\pi\xi)^{3/2}}N^{1/2}.
	\label{eqn:bdry_ring}
\end{equation}

Since longer chains experience higher entropy penalty, their reliable assembly 
requires somewhat stronger on-target binding, following a $N^{1/2}$ power law 
rather than $\ln N$ rule.

The third design rule is the \emph{concentration rule}, which establishes the finite interval of total 
building block concentrations that are optimally converted into the 
target structure. The upper and lower limits of this interval have  
different physical origins. At the upper limit, the absolute concentration of 
building blocks in the target structure $c_{targ}$ is not singular at $\mu^*$ 
(Fig.~\ref{fig:divergence}g-h). However, the total concentration of building 
blocks in all structures keeps growing, thus the conversion ratio drops off as 
$R\propto1/c_t$. At the lower limit, the conversion ratio is driven by 
$c_{targ}\propto e^{-1/\sqrt{c_t}}$ (see Appendix~\ref{app:statmech} for 
discussion of concentration 
units \cite{cates2015entropy}). Balancing the two 
effects results in the maximal conversion ratio at an intermediate optimal 
concentration:
\begin{equation}
	e^{-2}<\frac{c_t}{c_{opt}}<e^2;\quad c_{opt}=c_0\frac{N^3}{4}e^{\beta 
	v-S_b},
\end{equation}
where the lower and upper limits correspond to decrease of $R$ by a factor of 
$e$. The full width of the high-yield concentration range is $e^4\simeq 50$.

We test the three design rules by comparing the analytic limits with numerical 
computations of $R$ for a range of on-target bindings $\beta v$ and target 
structure sizes $N$ in Fig.~\ref{fig:rules}. As expected, significant 
conversion ratio is only achieved for parts of the $(N,\beta v)$ plane 
significantly below both the precision and boundary rules 
\eqref{eqn:prec_ring}--\eqref{eqn:bdry_ring}, with the second one providing a 
tighter bound (panel a). The concentration rule accurately estimates the 
location and width of the $R$ maximum (panel b). 

%alternative, short discussion
In this paper we investigate theoretically how to control the topology of 
structures self-assembled from heterogeneous building blocks, focusing 
on the central divergence in the partition function for both ordered and disordered binding 
energy matrices. By mitigating this divergence, we predict the quantitative 
boundary between designable and the intrinsically disordered regions of the 
design space \cite{williams2019dimers}. This framework opens the way to further 
quantitative investigations of formation of amyloids and design of structures 
of more complex topologies.

\section*{Acknowledgments}

We thank Agnese Curatolo, Chrisy Xiyu Du, Carl Goodrich, Ofer Kimchi, 
and Mobolaji Williams for useful discussions. The computational workflow in 
general and data management in particular for this work was primarily 
supported by the Signac data management framework 
\cite{signac_commat,signac_zenodo}. This research was funded by ONR 
N00014-17-1-3029. MPB is an investigator of the Simons Foundation.

\appendix

\section{Combinatorial theory}
\label{app:combin}
This section details the computation of the cluster partition function 
\eqref{eqn:Zcl} for our specific model system. We first introduce the 
hierarchical sum of the partition function over the topologies, lengths, and 
sequences. We then define the transfer matrix as the central element of the 
combinatorial calculation. We derive closed-form partition functions for linear 
chains and rings and indicate their convergence criteria. We show how the 
convergence and other properties of the partition function relate to the 
spectral properties of the transfer matrix. As we approach the divergence 
point, the chain and ring partition functions diverge at different rates, thus 
informing the entropic trade-off between them. We conclude the section with 
deriving the total concentrations and absolute yields of building blocks in 
closed form.

In order to start computing the partition functions, we need to look back at 
the hierarchical taxonomy of structures that can be assembled 
. The structure index $s$ so far did not refer to 
this hierarchy and was treated as one-dimensional. Without invalidating any of 
the summations in $s$, we can use it to refer to the topology, the length, and 
the specific sequence of the structure. In this case we can write the cluster 
partition function as a triple sum:
\begin{equation}
	\mathcal{Z}_{cl}=\sum\limits_\textsf{topology} \sum\limits_\textsf{length} 
	\sum\limits_\textsf{sequence} Q_s.
\end{equation}

The outermost sum is over the possible topologies. As we stated before, the 
monomers with two binding sites can only assemble two different topologies: 
chains and rings. For these topologies, we will write down two different 
expressions and add them up. The sum in length is relatively straightforward 
since the structures are one-dimensional, so the possible lengths are 
enumerated by a single set of natural numbers with some starting point. The sum 
in possible sequences is more tricky. However, since the energies of bonds and 
chemical potentials of the monomers are additive, the corresponding statistical 
weights are multiplicative and can be expressed via matrix products by 
following the approach of Ref.~\cite{murugan2015undesired}. The following 
sections derive the partition function expressions from the bottom up and 
develop the tools for their analysis.

\subsection{Chain and ring partition functions}
A linear chain consists of three components: the initial monomer driven only by 
the chemical potential, some number of subsequent monomers described by the 
transfer matrix, and the interaction of the last monomer with the solvent. The 
selection of the initial monomer is given by the vector of fugacities 
$\undervec{z}:z_i\equiv e^{\beta \mu^i}$. The number of intervening monomers 
varies between zero and infinity and is accounted by the sum in length. The 
terminus is taken as $\vec{1}$, a vector of all 1's since we assume that 
monomers don't have any specific interactions with the solvent. In this case, 
the partition function of all possible chains is given by:
\begin{equation}
	\mathcal{Z}_{chain}=\sum\limits_{n=0}^{\infty}\undervec{z}\mathbf{T}^n 
	\vec{1}.
\end{equation}

Note that for each term $n$, the chain has $n$ bonds that contribute energy and 
entropy and $n+1$ monomers that contribute the chemical potential. The 
contraction of the transfer matrix with the fugacities on the left and the 
terminus on the right can be taken outside of the summation due to linearity. 
The internal sum is then a simple geometric series in the matrix $\mathbf{T}$ 
that we will call the \emph{propagator} $\mathbf{D}$:
\begin{equation}
	\mathbf{D}\equiv\sum\limits_{n=0}^{\infty}\mathbf{T}^n=\left( 
	\mathbf{I}-\mathbf{T} \right)^{-1},
\end{equation}
where $\mathbf{I}$ is an $N\times N$ identity matrix. The geometric series does 
not always converge, and consequently the inverse does not always exist. This 
will have important physical implications below. For now, we can write down the 
chain partition function in a much more concise form:
\begin{equation}
	\mathcal{Z}_{chain}=\undervec{z}\mathbf{D} \vec{1}.
\end{equation}

By analogy with the chain, we can derive the ring partition function. The key 
difference between the chain and the ring is that there is no initial monomer 
and no terminus. Instead, the last monomer in the chain is bound to the first 
one, mathematically expressed with the trace operation $\Tr$. In order to 
account for this, we take the trace of the appropriate power 
of the transfer matrix. While the bending entropy of the bonds is already 
accounted for in the transfer matrix, the ring entropy penalty needs to be 
added manually. We also manually exclude the very short loops since they need 
excessively high bending. The resulting ring partition function is:
\begin{align}
	\mathcal{Z}_{ring}=&\sum\limits_{n=n_{min}}^{\infty}\frac{1}{(2\pi\xi 
		n)^{3/2}} \Tr(\mathbf{T}^n)\nonumber\\
	=&\frac{1}{(2\pi\xi)^{3/2}}\left( 
	\Tr(\Li_{3/2}(\mathbf{T}))-\sum\limits_{n=1}^{n_{min}-1}\frac{\Tr(\mathbf{T}^n)}{n^{3/2}}
	\right),
\end{align}
where $\Li_q$ is the polylogarithm function of order $q$ (here we use 
$q=3/2$).\cite{polylog} The polylogarithm is usually defined for a scalar 
argument, but we discuss a suitable generalization to matrix arguments 
below. Since the polylogarithm is defined via a sum starting from $n=1$, we 
need to manually subtract a finite number of terms of the sum. The 
polylogarithm remains the leading term that will govern the divergence 
properties of the ring partition function, described below.

The total partition function of all possible structures in this system is the 
sum over the two possible topologies:
\begin{equation}
	\mathcal{Z}_{cl}=\mathcal{Z}_{chain}+\mathcal{Z}_{ring}.
\end{equation}

This cluster partition function is the basis for computing the total monomer 
density and absolute yields of target structures. However, to predict those 
quantities it is crucially important to understand the properties of the 
divergence of both the geometric series and the polylogarithm. Those 
divergences, in turn, are directly related to the spectral properties of the 
transfer matrix, as discussed below.
\subsection{Spectral analysis}
Both the propagator and the polylogarithm are defined as sums of series in 
powers of the transfer matrix. In order to clarify the conditions of 
convergence of this series, we can analyze it in terms of the spectrum of the 
transfer matrix. The eigendecomposition of the transfer matrix can be written 
down as following:
\begin{equation}
	\mathbf{T}=\mathbf{P}\bm{\Lambda}\mathbf{P}^{-1},
\end{equation}
where $\mathbf{P}$ is a transformation matrix of eigenvectors and 
$\bm{\Lambda}$ is a diagonal matrix of eigenvalues. Usually, the matrix 
$\mathbf{P}$ will be unitary, since the eigenvectors are typically orthogonal 
to each other for most matrices that occur in physics. However, in the present 
case we are dealing with a non-symmetric matrix $\mathbf{T}$. While the 
spectrum can be found even for a non-symmetric matrix, the orthogonality of 
eigenvectors is not guaranteed. Because of that, the inverse of the transform 
matrix does not necessarily coincide with the conjugate transpose 
$\mathbf{P}^{-1}\neq \mathbf{P}^*$. The eigenvectors and eigenvalues together 
form the spectrum of the matrix, which for small matrices can be easily 
computed numerically once and looked up after that.

The properties of 
$\mathbf{T}$ impose some constraints on its spectrum. If none of the binding 
energies are positive-infinite, the matrix consists of positive elements only, 
which by Perron-Frobenius theorem guarantees that the top eigenvalue 
$\lambda_0$ is real, positive, and an all-positive eigenvector can be found for 
it. However, since the transfer matrix $\mathbf{T}$ depends on the binding 
energy matrix $\mathbf{E}$, which is not symmetric, there is no guarantee that 
the rest of the spectrum is real, only that it lies within a complex circle of 
radius $\lambda_0$.

In terms of the spectrum, the propagator and the polylogarithm can be computed 
as following:
\begin{align}
	\mathbf{D}=&\mathbf{P} 
	\left(\mathbf{I}-\bm{\Lambda}\right)^{-1}\mathbf{P}^{-1}\\
	\Li(\mathbf{T})=&\mathbf{P}\Li(\bm{\Lambda})\mathbf{P}^{-1},
\end{align}
where the inverse and the polylogarithm of the diagonal $\mathbf{\Lambda}$ 
matrix can be computed elementwise. The convergence of both of them is then 
dependent on the top eigenvalue. Importantly, the convergence condition 
$\lambda_0<1$ is the same for both functions. Since $\lambda_0$ depends on the 
components of the transfer matrix, then as these components vary, both the 
chain and the ring partition functions will diverge at exactly the same point, 
though at different rates which we analyze below.

While the top eigenvalue is sufficient to determine whether or not the sum over 
all statistical weights diverges, the structure of that sum also depends on all 
other eigenvalues. A simple way to see that is to examine the sequence of 
weights of the rings of variable size. The weights are proportional to the 
traces of powers of the transfer matrix, which in turn are related to sums over 
eigenvalues:
\begin{equation}
	Q_n\propto \Tr(\mathbf{T}^n)=\sum_{k=0}^{N-1} \lambda_k^n,
\end{equation}
where $N$ is the number of monomers, equal to the dimension of the matrix 
$\mathbf{T}$ and thus the number of eigenvalues. Even though the eigenvalues 
themselves can be complex, the sum over them is always real, resulting in a 
real $Q_n$. However, for certain values of ring length $n$ all of the 
eigenvalues can add up in phase and give a boost to the statistical weight. 
Such a boost will be an evidence of the successful programming of a target 
structure into the energy matrix, as shown in the Topological Design Rules 
section below. A similar spectral analysis can be carried out for the assembly 
of linear chains.

The transfer matrix spectrum is useful for us in two ways: while the lead 
eigenvalue governs the overall divergence of the partition functions, all other 
eigenvalues determine whether specific target 
structures are robustly formed close to that divergence. Therefore, the 
spectrum is a \emph{collective metric} that quantifies the properties of a 
given set of building blocks. In order to design an optimal set of building 
blocks for self-assembly, we can state the desired spectral properties of the 
transfer matrix and then attempt to pick the blocks that can realize those 
properties.

\subsection{Divergence analysis}
The divergence properties of the partition function contain important knowledge 
about the design limitations. In order to elucidate these limitations, we need 
to look at the asymptotic behavior of the partition functions. Generically, as 
we approach the divergence in the space of chemical potentials, the lead 
eigenvalue of the transfer matrix will have the following form:
\begin{equation}
	\lambda_0 = \exp(\beta(\mu - \mu^*)),
\end{equation}
where the divergence point $\mu^*$ depends in a complicated way on the 
properties of the binding energies. Exactly at the divergence point 
$\lambda_0=\exp(0)=1$. Since the divergence point is the same 
for the linear chains and the closed rings, we can compare the asymptotic forms 
of the two partition functions. For the chain partition function, the 
asymptotic form is quite simple:
\begin{equation}
	\mathcal{Z}_{chain}=\undervec{z} (\mathbf{I}-\mathbf{T})^{-1} \vec{1} 
	\propto A(1-\lambda_0)^{-1} \propto A\left(\mu^*-\mu \right)^{-1},
\end{equation}
where $A$ is a non-singular positive constant.

For the ring partition function, the analysis is slightly more complicated and 
has to do with the properties of the polylogarithm function. As the argument 
$x\equiv e^w$ of the  polylogarithm $\Li_q(x)$ approaches $x=1$, the singular 
term can be separated from the analytic term:\cite{polylog}
\begin{equation}
	\Li_q(e^w)=\Gamma(1-q)(-w)^{q-1}+\order{w^0},
\end{equation}
where $\order{w^0}$ is a series in non-negative powers of $w$ that does not 
affect the singularity. In our case $q=3/2$, so the singular part is 
$(-w)^{1/2}$. Note that the Gamma function $\Gamma(-1/2)$ is negative, so that 
the singular part of the ring partition function is negative. We can therefore 
write it down as following:
\begin{equation}
	\mathcal{Z}_{ring}\propto -B\left(\mu^* -\mu\right)^{1/2}+C,
\end{equation}
where $B,C$ are non-singular positive constants. Note that as $\mu \to \mu^*$, 
the ring partition function itself does not approach infinite value, even 
though it is not defined for $\mu> \mu^*$.

However, the interesting part of the divergence is the behavior of total 
concentrations, i.e. the derivative of the cluster partition function 
$\mathcal{Z}_{cl}= \mathcal{Z}_{chain}+\mathcal{Z}_{ring}$. The derivative 
shifts the divergence exponents by 1, giving:
\begin{equation}
	c_t\propto \pdv{\mathcal{Z}_{cl}}{\mu}=
	\underbrace{A(\mu^* -\mu)^{-2}}_\text{chain} + 
	\underbrace{B(\mu^* -\mu)^{-1/2}}_\text{ring}.
	\label{eqn:div_comp}
\end{equation}

Each of the two divergent terms corresponds to the number of building blocks 
bound up in either chains or rings. We are interested in which of these terms 
if larger. At any finite density, $(\mu^*-\mu)>0$, and the relative balance of 
the two terms depends on the values of $A,B$, which in turn depend on all of 
the design space variables in a complicated way. Figuring out this complicated 
dependence is the goal of design. A system well designed for assembly of chains 
will have the term with $A$ dominate, whereas a system well designed for 
assembly of rings will have the term with $B$ dominate. However, the terms 
still diverge at different rates. Regardless of our design efforts, for very 
high concentrations $c_t$ we will get arbitrarily close to $\mu^*$ 
and the assembly of chains will always dominate. We use this fact below in the 
derivation of the concentration design rule.

It is important to note that these very high concentrations $c_t$ 
are not always possible, since the building blocks will either reach 
close-packing density, or different clusters will be so close to each other 
that ignoring their interactions will be impossible. In either case, the 
cluster-based theory is likely to break down in that limit.

There is two more qualitative features that can be extracted from the analysis 
of this divergence. First, close to the divergence point 
Eqn.~\eqref{eqn:div_comp} becomes a simplified form of an equation of state 
that connects chemical potential to the concentration. For example, for a 
system dominated by chains, we can easily solve for the chemical potential in 
terms of the experimentally controllable concentration:
\begin{equation}
	\mu \simeq \mu^* - \left( \frac{c_t}{A} \right)^{-1/2}.
\end{equation}

This approximate value of chemical potential can then be plugged into any other 
$\mu$-dependent expressions in the grand canonical theory. For example, the 
lead eigenvalue of the transfer matrix is then $\lambda_0\simeq 
1-(c_t/A)^{-1/2}$, asymptotically approaching 1 as expected.

The second qualitative feature is the asymptotic behavior of the conversion 
ratio at large concentrations. The partition function diverges because it is a 
sum over an \emph{infinite number} of statistical weights $Q_s$. Each 
statistical weight is an analytical function of chemical potential $\mu$, thus 
it does not diverge at $\mu^*$. Similarly, if the target structure aggregates a 
\emph{finite} number of structures $s$, its statistical weight does not diverge 
either. Therefore, the asymptotic behavior of the conversion ratio is governed 
exclusively by the divergence of the total concentration:
\begin{equation}
	R_{targ}=\frac{1}{\beta c_t}\pdv{Q_{target}}{\mu}\propto 
	\frac{1}{c_t}.
\end{equation}

In other words, optimal conversion of raw monomers into target structures will 
always be achieved at finite building block concentration. Arbitrarily large 
building block concentrations favor formation of larger and larger structures, 
thus they will always suppress the formation of any finite target structure.

\subsection{Computing yields}
In order to complete the theory, we need to predict the absolute yields and 
total concentrations of structures of interest. Above we showed the general 
formula for total concentration in vector derivative format. In order to 
perform that derivative analytically, it will be more convenient to rewrite the 
expression in index notation:
\begin{equation}
	c_{t,l}=c_0\frac{1}{\beta}\pdv{\mathcal{Z}_{cl}}{\mu^l},
\end{equation}
where we highlight that a derivative of a scalar ($\mathcal{Z}_{cl}$) with 
respect to a column vector ($\mu^l$) is a row vector ($c_{t,l}$). 
Similarly, the vector derivative of any higher-rank object adds an extra index 
to the result. For an example, an important building block of our theory is the 
transfer matrix $\mathbf{T}$ defined elementwise by Eqn.(2) of main text. Its 
derivative is:
\begin{equation}
	\frac{1}{\beta}\pdv{T^i_{\;j}}{\mu^l}=T^i_{\;j}\delta_{jl},
\end{equation}
where there is no summation in repeated index $j$. Since each index might be 
repeated more than twice and not always summed over, throughout this section we 
do not assume the Einstein convention and instead write the summations out 
explicitly.

In order to find the derivative of the propagator, we write down the identity 
that defines it:
\begin{equation}
	\sum\limits_{j}D^m_{\;j}\left( \delta^j_{\;r}-T^j_{\;r} 
	\right)=\delta^m_{\;r}.
\end{equation}

We take $\mu^l$ derivatives on both sides, multiply on the right with the 
propagator $D^r_{\;k}$, and perform the sum in $r$ to get:
\begin{equation}
	\frac{1}{\beta} \pdv{D^m_{\;k}}{\mu^l}=\sum\limits_{i,j}D^m_{\;i} T^i_{\;j} 
	\delta_{jl} D^j_{\;k} = \left(\sum\limits_{i} D^m_{\;i} 
	T^i_{\;l}\right)D^l_{\;k}.
\end{equation}

The propagator derivative allows us to write down the derivative of the chain 
partition function:
\begin{align}
	&c_{chain,l}=\frac{c_0}{\beta}\pdv{\mathcal{Z}_{chain}}{\mu^l}=
	\frac{c_0}{\beta} \pdv{\mu^l}\left(\sum\limits_{i,j}z_i D^i_{\;j} 
	1^j\right)\nonumber\\
	=&c_0 \left[ \left( \sum\limits_{j} z_l D^l_{\;j}1^j \right) +
	\left( \sum\limits_{i,j} z_i D^i_{\;j}T^j_{\;l} \right) \left( 
	\sum\limits_{j}  D^l_{\;j}1^j \right) \right].
	\label{eqn:c_chain}
\end{align}

While the expression above appears cumbersome, it amounts to several lookups 
and multiplications of matrices $\mathbf{T},\mathbf{D}$ that are already known. 
In a similar way, the main component of the ring partition function is the 
trace of a power of the transfer matrix $\Tr(\mathbf{T}^n)$. Its derivative can 
be shown to be:
\begin{equation}
	\frac{1}{\beta}\pdv{\mu^l}\Tr(\mathbf{T}^n)=n (\mathbf{T}^n)^l_{\;l},
\end{equation}
so that the derivative is read off the diagonal of a matrix power (a trace will 
have summed up that diagonal). This expression, up to a prefactor, gives the 
concentration of building blocks bound up in rings of a specific size $n$. By 
summing such expressions, we can express the total concentration of building 
blocks bound up in rings of all sizes:
\begin{align}
	c_{ring,l}=&\frac{c_0}{\beta}\pdv{\mathcal{Z}_{ring}}{\mu^l}\nonumber\\
	=&\frac{c_0}{(2\pi \xi)^{3/2}}\left[ 
	\left(\Li_{1/2}(\mathbf{T})\right)^l_{\;l} - 
	\sum\limits_{n=1}^{n_{min}-1}n^{-1/2}(\mathbf{T}^n)^l_{\;l} \right].
	\label{eqn:c_ring}
\end{align}

Note that the order of the polylogarithm and the power of $n$ in the second sum 
both shifted by 1. The polylogarithm of any order diverges at the same value of 
the argument, and thus the polylogarithm of a matrix can be evaluated via the 
same spectral method.

It is important to note that the equations \eqref{eqn:c_chain} and 
\eqref{eqn:c_ring} are valid for any transfer matrix regardless of the 
symmetries or design decisions built in. We are still at liberty of assigning 
the binding energies $E^i_{\;j}$, the chemical potentials $\mu^l$, and the 
bending rigidity $P$. The total concentration and related yield expressions 
then directly connect the points in design space (encoded in $\mathbf{T}$) with 
the self-assembly outcomes (encoded in $c_{chain,l}$ and $c_{ring,l}$).

The full expressions \eqref{eqn:c_chain} and \eqref{eqn:c_ring} are also useful 
for studying the difference in concentrations of different building block 
types. Varying the concentrations, or chemical potentials, has been shown to 
increase the yields of linear chains in Ref.~\cite{murugan2015undesired}. 
However, in the present paper we are mainly interested in assembling closed 
rings, which does not require variation of chemical potential. Therefore, we 
can compute the total concentrations of building blocks of all types in chains 
and rings by summing the above expressions in $l$. The resulting expressions 
take much more concise forms and can be written down in matrix notation:
\begin{align}
	c_{chain}=&\sum\limits_{l}c_{chain,l}=c_0\left( 
	\undervec{z}\left(\mathbf{D}+\mathbf{D}\mathbf{T}\mathbf{D}\right)\vec{1} 
	\right)\label{eqn:cchain}\\
	c_{ring}=&\sum\limits_{l}c_{ring,l}\nonumber\\
	=&\frac{c_0}{(2\pi \xi)^{3/2}}\left[ 
	\Tr\left(\Li_{1/2}(\mathbf{T})\right) - 
	\sum\limits_{n=1}^{n_{min}-1}n^{-1/2}\Tr(\mathbf{T}^n) \right]
\end{align}

We use these fully analytic expressions for arbitrary $\mathbf{T}$ to compute 
the self-assembly yields for the three scenarios in Fig.~3 of main text. Below 
we assume a specific functional form of the transfer matrix and derive the 
three design rules analytically.

\section{Topological design rules}
\label{app:desrules}
In this section we show derivations of the topological design rules for optimal 
assembly of closed rings. First, we introduce a special form of the binding 
energy matrix that can be diagonalized in closed from. Then, we show how 
special properties of the matrix spectrum are used to derive the three design 
rules: precision, boundary, and concentration. Lastly, we comment on several 
aspects of numerical computations of the conversion ratio.

\subsection{Analytic diagonalization}
In order to preferentially assemble closed rings, we consider a binding energy 
matrix that is shift-periodic:
\begin{equation}
	\begin{cases}
		E^i_{\;j}=v,&j=i+1\\
		E^i_{\;j}=\epsilon,&j\neq i+1
	\end{cases},
\end{equation}
where the equality of indices is taken modulo $N$. We also assume that all 
chemical potentials are the same $\mu^i=\mu$ and there the bending entropy is
$S_b$. The resulting transfer matrix $\mathbf{T}$ is not symmetric, and so it 
doesn't fulfill the most commonly used sufficient condition of being 
diagonalizable with orthogonal eigenvectors. However, the shift-periodicity is 
also sufficient for diagonalizing. We prove this by construction by considering 
the eigenvectors of the Fourier form, so that the $j$'th element of $k$'th 
eigenvector is:
\begin{equation}
	v^{j,(k)}=\frac{1}{\sqrt{N}}e^{i2\pi\frac{jk}{N}}.
\end{equation}

A direct computation verifies this ansatz to be the eigenvectors corresponding 
to complex eigenvalues:
\begin{align}
	\lambda_k =& \lambda e^{i2\pi \frac{k}{N}}+\Delta 
	\delta_{k,0}\label{eqn:lambdak}\\
	\lambda=&e^{\beta \mu}e^{S_b}\left( e^{-\beta v}-e^{-\beta \epsilon} 
	\right)\\
	\Delta=&e^{\beta \mu}e^{S_b} N e^{-\beta \epsilon},
\end{align}
where the top eigenvalue $\lambda_0=\lambda+\Delta$ is real as expected by 
Perron-Frobenius theorem. In order to ensure convergence, we assume 
$\lambda+\Delta<1$. Additionally, the top eigenvalue can be written as 
$\lambda_0=\exp(\beta(\mu-\mu^*))$, which defines the divergence point $\mu^*$ 
as:
\begin{equation}
	\beta \mu^*=-S_b-\ln(e^{-\beta v}+(N-1)e^{-\beta \epsilon}).
\end{equation}

Below we use several properties of the above spectrum to derive three 
analytical rules that ensure robust assembly of closed rings of target length 
$N$.

\subsection{Precision rule}
First, we need to ensure that when rings are assembled, they preferentially 
have size $N$. Consider the sequence of statistical weight of rings of sizes 
$n$:
\begin{equation}
	Q_n=\frac{1}{(2\pi\xi n)^{3/2}}\Tr(\mathbf{T}^n)=\frac{1}{(2\pi\xi 
		n)^{3/2}} \sum\limits_{k=0}^{N-1}\lambda_k^n.
\end{equation}

Since all eigenvalues $\lambda_{k\neq 0}<\lambda_0<1$, this is a sequence that 
overall decays with $n$, but with a particular structure. By plugging in the 
analytical diagonalization \eqref{eqn:lambdak}, we get the following expression:
\begin{equation}
	\Tr(\mathbf{T}^n)\simeq (\lambda+\Delta)^n+\lambda^n n \delta(n,Nr),
\end{equation}
where $r$ is an integer counting how many times the full sequence has repeated. 
In other words, the statistical weight gets a boost every $N$ elements. Is this 
boost significant? In other words, does it significantly affect the structures 
assembled? This boost will be significant if the second term in the sum is much 
larger than the first one for the weight of the target structure $n=N$:
\begin{align}
	N\lambda^N \gg& (\lambda+\Delta)^N\label{eqn:boost}\\
	\ln N \gg& N \ln(1+\Delta/\lambda)\\
	\frac{\ln N}{N} \gg& N \frac{e^{-\beta \epsilon}}{e^{-\beta v}}\\
	\frac{N^2}{\ln N} \ll& e^{-\beta (v-\epsilon)}.\label{eqn:rule1}
\end{align}

From the spectral point of view, the boost is significant when the eigengap 
$\Delta$ between the leading and subleading eigenvalues is sufficiently small. 
In this case the nearly-divergent series of $Q_n$ has a nontrivial structure, 
different from the simple geometric series $\lambda_0^n$.

From the physical point of view, the boost is significant when the binding 
energy \emph{contrast} ($v-\epsilon$) between on-target and off-target binding 
is sufficiently strong to overcome the entropy of many possible off-target 
bindings. This contrast is typically limited by the properties of the platform 
that realizes the specific interactions. The derived limit is remarkably 
similar to the ``$\log n$ limit'' of Ref.~\cite{murugan2015undesired}.

\subsection{Boundary rule}
Having established the precision rule \eqref{eqn:rule1} and assuming that it is 
fulfilled, we can now focus on the boundary rule. We now know that of all 
rings, those of length $N$ would preferentially assemble. However, would 
structures of length $N$ want to be closed rings or linear chains? We can 
answer that by comparing the concentration of particles bound in chains and in 
rings, both of length exactly $N$:
\begin{equation}
	c_{ring}^{N}\overset{?}{\gg}c_{chain}^{N}.
\end{equation}

The concentration of building blocks in rings can be conveniently expressed via 
the trace:
\begin{equation}
	c_{ring}^{N}=\frac{c_0}{(2\pi\xi)^{3/2}}N^{-1/2}\Tr(\mathbf{T}^N)\simeq
	\frac{c_0}{(2\pi\xi)^{3/2}}N^{-1/2} N \lambda^N,
	\label{eqn:crN}
\end{equation}
where we used the condition of boost \eqref{eqn:boost}.

The concentration of building blocks in chains can be expressed in linear 
algebra form:
\begin{equation}
	c_{chain}^{N}=N \undervec{z}\mathbf{T}^{N-1}\vec{1}.
\end{equation}

We can make further progress on this expression by using the orthogonality of 
eigenvectors that results from the special form of the binding energy matrix. 
First, notice that the column vector $\vec{1}$ and the row vector 
$\undervec{z}$ are proportional to the top eigenvector and its dual, 
respectively:
\begin{align}
	\vec{1}=&\sqrt{N}\vec{v}^{0}\\
	\undervec{z}=&e^{\beta \mu}\sqrt{N}\undervec{v}_{(0)}\\
	\undervec{v}_{(0)}&\mathbf{T}^n \vec{v}^{(0)}=\lambda_0^n.
\end{align}

Second, note that $e^{\beta \mu}=e^{\beta \mu^*}\lambda_0$. By using both of 
these facts, we find that the concentration of building blocks in chains only 
depends on the top eigenvalue:
\begin{equation}
	c_{chain}^{N}=N^2 e^{\beta\mu^*} \lambda_0^N.
	\label{eqn:ccN}
\end{equation}

Comparing the expressions \eqref{eqn:ccN} and \eqref{eqn:crN}, we derive the 
following criterion:
\begin{align}
	c_{chain}^N\ll&c_{ring}^N\\
	e^{\beta\mu}\ll& 
	\frac{1}{(2\pi\xi)^{3/2}}N^{-3/2}\left(1+\frac{\Delta}{\lambda}\right)^{-N}\\
	\beta v\ll& S_b-\frac{3}{2}\ln(2\pi\xi)-\frac{3}{2}\ln N-N^2 
	e^{-\beta\epsilon}e^{\beta v}.\label{eqn:rule2p}
\end{align} 

The last term in the above expression is quite small because of the precision 
rule \eqref{eqn:rule1}. To the lowest order, the term can be dropped 
completely. However, we can make a slightly more precise statement by 
recognizing that the expression \eqref{eqn:rule2p} has a self-consistent form 
$\beta v\ll f(\beta v)$. We can get a better approximation to this limit by 
plugging the whole right hand side as an argument into the right hand side 
$\beta v\ll f(f(\beta v))$, resulting in:
\begin{equation}
	\beta v\ll S_b-\frac{3}{2}\ln(2\pi\xi)-\frac{3}{2}\ln N- \frac{e^{-\beta 
			\epsilon}}{(2\pi\xi)^{3/2}} N^{1/2} e^{-N^2 e^{-\beta 
			\epsilon}e^{\beta v}}.
\end{equation}

Because of the precision rule, the argument of the exponent $e^{-N^2 e^{-\beta 
		\epsilon}e^{\beta v}}$ is now very small, making it approximately $1$ 
		and 
resulting in Eqn.~10 of the main text.

Physically, this limit implies that the extra bond of the closed ring $e^{\beta 
	v-S_b}$ has to be stronger than the ring entropy penalty $1/(2\pi\xi 
	N)^{3/2}$.

\subsection{Concentration rule}
Having established the energetic constraints on precision of structures, we 
need to complement them with concentration constraints. In deriving the first 
two rules, we compared the concentrations of building blocks in rings of 
different lengths, and in chains of length $N$. At the same time, the 
divergence analysis suggests that most of building blocks would still be in 
long linear chains $c_t\simeq c_{chain}$. This implies that linear chains 
constitute the bulk of total concentration and are responsible for selecting 
the chemical potential. We will use this fact to derive the $R$ for 
\emph{rings} by assuming that $c_t$ is dominated by \emph{chains}.

The total concentration of building blocks in all chains is given by 
\eqref{eqn:cchain}. However, the expression can be further simplified because 
of orthogonality of eigenvectors:
\begin{align}
	&c_{chain}=c_0\left(\undervec{z}\left(\mathbf{D}+
	\mathbf{D}\mathbf{T}\mathbf{D}\right)\vec{1}\right)\nonumber\\
	=&c_0 N e^{\beta\mu^*}\lambda_0 
	\left(\frac{1}{1-\lambda_0}+\frac{\lambda_0}{(1-\lambda_0)^2}\right)
	=\frac{c_0 N e^{\beta\mu^*}\lambda_0}{(1-\lambda_0)^2}.
\end{align}

We plug in $\lambda_0=\exp(\beta (\mu-\mu^*))$ and consider the near-divergent 
regime:
\begin{align}
	c_t\simeq c_{chain}\simeq& \frac{c_0 N e^{\beta v-S_b}}{\left(\beta 
		(\mu^*-\mu)\right)^2}\\
	\beta (\mu^*-\mu)\simeq&\left(\frac{c_t}{c_0 N K}\right)^{-1/2},
\end{align}
where we denoted $K\equiv e^{\beta v-S_b}$. The conversion ratio of monomers 
into rings of length $N$ is:
\begin{equation}
	R=\frac{c_{ring}}{c_t}=\frac{c_0}{c_t}\frac{N^{1/2}}{(2\pi\xi)^{3/2}} 
	e^{N\beta (\mu-\mu^*)}.
\end{equation}

By plugging in the expression for the chemical potential and denoting 
$c_t/c_0\equiv X$, we get:
\begin{equation}
	R=\frac{N^{1/2}}{(2\pi\xi)^{3/2}}\frac{1}{X}e^{-N^{3/2}(X/K)^{-1/2}},
\end{equation}
which expresses the asymptotic shapes of the $R$ graph tails shown in Fig.~4b 
of main text. In order to find the location of the $R$ maximum, we simply set 
the $X$ derivative of $\ln R$ to zero:
\begin{align}
	\ln R=&const-\ln X-N^{3/2}K^{1/2}X^{-1/2}\\
	\frac{d\ln R}{dX}=&0\quad\Rightarrow\quad X_{opt}=\frac{N^3 
		K}{4}=\frac{N^3}{4}e^{\beta v-S_b}.
\end{align}

The value of $X_{opt}$ sets the location of the optimal concentration for 
assembly $c_{opt}=c_0 X_{opt}$. In order to establish the width of the optimal 
concentration interval, we note that the graph of $R(X)$ looks approximately 
parabolic in log-log coordinates. Let $X=X_{opt} e^A$ for some unknown small 
$A$. By expanding $\ln R$ up to quadratic order in $A$, we get:
\begin{equation}
	\ln R\approx \ln R_{max}-\frac{A^2}{4}+\order{A^3}.
\end{equation}

The width of the concentration interval is given by the decay of the conversion 
ratio by a factor of $e$, corresponding to $\ln R-\ln R_{max}=-1$. This is 
achieved for $A=\pm 2$, which specifies the width of the near-optimal 
concentration interval on the logarithmic axis. On linear axis, this 
corresponds to the concentration belonging to the range $c_t\in 
[c_{opt}e^{-2},c_{opt}e^2]$, which has a width of about a factor of $e^4\approx 
50$.

Note that the optimal concentration $c_{opt}$ is proportional to the reference 
concentration and thus seemingly depends on the concentration convention 
chosen. However, it also includes the bond free energy $v$, which precisely 
cancels the dependence on the convention. For any conversion we might choose, 
the predicted optimal concentration would be the same.

\section{Statistical mechanics for self-assembly}
\label{app:statmech}
This section connects the key tenets of statistical mechanics in application to 
self-assembly of classical particles. First, we establish the notational 
conventions on measurement of concentrations. Second, we explain the difference 
between two versions of partition functions used in the literature. Third, we 
introduce the metrics of self-assembly yield we use in the paper. Fourth and 
last, we introduce several ideas from polymer theory to compute the 
conformational entropy of linear chains and closed rings.

\subsection{Concentration convention}
In this paper we deal with interactions of fully classical building blocks, 
which requires a careful re-defining of the units for measuring their volume 
concentration. Statistical mechanics was originally created to treat atomic and 
molecular systems that were soon discovered to obey quantum mechanics; it is 
important to not carry over the quantum intuitions to a fully classical system, 
as discussed in detail in Ref.~\cite{cates2015entropy}. Specifically, the phase 
space measure $h$ is intimately tied to the units of concentration and 
classical bond entropies, but in the classical case it is not tied to Planck's 
constant.

We illustrate the question of concentration in this section by considering a 
system of one building block and a system of two building blocks; further 
sections will develop a combinatorial theory to account for 
larger structures and their copy numbers. First, consider a trivial system of 
one particle in a large box of volume $V$. The statistical weight of all 
configurations of such a particle is:
\begin{equation}
	\mathcal{Z}_1=e^{\beta\mu}\int\limits_{\vec{x},\vec{p}}\frac{d\vec{x}d\vec{p}}{h^3}
	=e^{\beta\mu}\frac{V}{\lambda_{th}^3}=e^{\beta\mu}c_0 V,
\end{equation}
where $h$ is a measure of phase space. It has the same dimensions $ML^2/T$ as 
Planck's constant but its value is arbitrary. Integrating out the momentum of 
the particle we are left with the factor of thermal wavelength $\lambda_{th}$, 
which we use to define the \emph{reference concentration} $c_0\equiv 
\lambda_{th}^{-3}$. Since $h$, $\lambda_{th}$, and $c_0$ are rigidly linked to 
each other, we are still left with one undetermined constant.

Now consider a system of two particles that interact via a finite-range 
potential $u(\vec{\Delta x})$ such that $u\to 0$ for $\vec{\Delta x}\to 
\infty$. For simplicity, we only consider their relative 
position and ignore the orientation and the combinatorial factors associated 
with distinguishability.  The partition function of all configurations of these 
two particles can be computed by first placing one particle anywhere in the 
volume, and then placing the other particle in all possible relative positions. 
Since the particles now interact, we add the Boltzmann factor associated with 
their interaction energy:
\begin{equation}
	\mathcal{Z}_2=e^{2\beta\mu}\int\frac{d\vec{x}}{\lambda_{th}^3}
	\frac{d\vec{\Delta x}}{\lambda_{th}^3}
	e^{-\beta u(\Delta \vec{x})}.
\end{equation}

We break down the potential into two terms: its lowest value and deviation from 
it: $u(\vec{\Delta x})=u_{min}+\Delta u(\vec{\Delta x})$. The value of $u$ is 
only significant in some small, but finite 
interaction volume $V_{int}$. Therefore the integration in $\vec{\Delta x}$ 
breaks into two regions: small distance where the interaction is important, and 
large distance where the two particles are effectively independent:
\begin{align}
	\mathcal{Z}_2=&\frac{e^{2\beta\mu}}{(\lambda_{th}^3)^2}\left( V 
	V_{int}e^{-\beta u_{min}}+V^2 \right)\\
	=&\left(c_0 V e^{\beta\mu}\right)\left(c_0 V_{int}e^{-\beta 
		u_{min}}e^{\beta\mu}\right)+
	\left(c_0 V e^{\beta\mu}\right)^2.\label{eqn:Z2}
\end{align}

Here, the first term describes two particles classically bound to each other, 
the second term describes two particles that are independent. It seems that the 
expression closely depends on the yet-undetermined value of $c_0$. However, 
this value would cancel out from thermodynamic observables. For instance, the 
odds of observing a bound pair of particles (a dimer) as opposed to two 
separate particles is the ratio of the two terms, independent of our choice of 
$c_0$.

In development of a more complex theory, we want to only count the structures 
where all particles are connected by bonds. Each of such structures starts with 
a first particle that can explore the whole box volume $V$, and then other 
particles are added to it. Adding another particle to the chain should amount 
to just ``paying'' the costs of creating it $\mu$ and bonding it to the 
previous one in the chain $v$. By examining the expression \eqref{eqn:Z2}, we 
define:
\begin{align}
	c_0 V_{int} e^{-\beta u_{min}}\equiv& e^{-\beta v}\\
	v=&u_{min}-\frac{1}{\beta}\ln(c_0 V_{int}),
\end{align}
where $u_{min}$ is the \emph{bond energy} (minimum of the potential) and $v$ is 
the \emph{bond free energy}. The difference between the two 
$\frac{1}{\beta}\ln(c_0 V_{min})$ is the classical bond entropy that does not 
have a quantum mechanical analog. This correction cannot be simply ignored. For 
example, if we use the real Planck's constant $h$ for the phase space measure, 
the thermal wavelength $\lambda_{th}$ would become really small because the 
mass of colloids is much larger than the mass of atoms. This would make $c_0$ a 
very large number, and the bond entropy would be much larger than bond energy. 
However, whether or not two particles bind to each other would depend on the 
\emph{competition} between the bond entropy and the translational entropy of 
a unbound particle $\frac{1}{\beta}\ln(c_0 V)$. This competition is 
unaffected by our choices of units, therefore the choice should be governed by 
computational convenience.

In the derivations that follow, we leave the expressions in terms of the 
reference concentration $c_0$ and the bond [free] energy $v$ that absorbed the 
entropy term. We denote off-target binding energy by $\epsilon$ and assume that 
the entropy of off-target bonds is the same as that of on-target. This allows 
us to proceed with combinatorial calculations. At the end, we make predictions 
about optimal concentrations that need to be converted into real units. We 
propose two different strategies for unit conversion that exploit the freedom 
of choice in different ways but give the same numerical predictions.

Under the first strategy, we set the desired measurement unit $\lambda_{th}$, 
which fixes the units of concentration $c_0=\lambda_{th}^{-3}$. Then we compute 
the entropy correction $\ln(c_0 V_{min})$ and subtract it from each bond 
energy. This correction depends on the interaction volume, and thus on the 
interaction potential shape of the building blocks. The resulting values of 
$c_0,v,\epsilon$ can be substituted into the yield formulas.

Under the second strategy, we nullify the entropy correction by setting 
$c_0=1/V_{int}$, which makes $\ln(c_0 V_{min})=0$. The bond free energies are 
then equal to the bond energies, but concentration units are related to the 
interaction volume.

We return to discussion of concentration convention in context of the 
concentration design rule to show that it is independent of our choices of 
conventions.

Lastly, we assume throughout the paper that the concentrations are always in 
the dilute limit, that is, we can treat particle clusters as independent from 
each other. There is a hard upper limit on number concentration of building 
blocks $c_{max}=a^{-3}$ set by particle size $a$ at close packing. Close to 
that concentration, we expect the dilute limit predictions to break down as the 
system of building blocks starts to vitrify.

\subsection{Cluster and box partition functions}
The textbook definition of the grand canonical ensemble prescribes summing over 
all possible configurations of particles in the system, including the number of 
particles. This prescription can be interpreted in two ways: either focusing on 
all \emph{unique} particle configurations, or on \emph{all possible} 
configurations, including their copy numbers. We show below that these two 
summations are both necessary and closely related to each other.

The first form of summation is what we call a \emph{cluster partition 
	function}, and is a direct sum of statistical weights of all \emph{unique} 
clusters, 
without taking their copy numbers into account:
\begin{equation}
	\mathcal{Z}_{cl}=\sum\limits_{s}Q_s.
	\label{eqn:Zcl}
\end{equation}

Note that the cluster partition function is somehow ignorant of two factors: 
the box in which the clusters can move, and the copy number of identical 
clusters. Accounting for the volume of the box is the same 
thing as accounting for the translational entropy of the cluster. Taking into 
account the volume of the box adds a factor of $V$, while integrating out the 
kinetic energy of motion of the whole cluster adds a factor of reference 
concentration, related to the thermal wavelength $c_0=\lambda_{th}^{-3}$. Note 
that the reference concentration only appears once, regardless of the number of 
particles in the cluster, since all other factors of thermal wavelength are 
absorbed in the calculation of vibrational entropy, as explained in the 
previous section following Ref.~\cite{cates2015entropy}.

Apart from the translational entropy, we need to account for the copy number 
$m_s$ of each type of cluster: while clusters of the same type $s$ are 
indistinguishable from each other, they can be counted. Each cluster type 
appears between zero and infinity times within the box, regulated by the 
corresponding statistical weight. Performing the sum over all cluster types and 
their copy numbers, we get the \emph{box partition function}:
\begin{align}
	\mathcal{Z}_{box}=&\sum_{\{m_s\}}\prod\limits_{s}\frac{1}{m_s!}\left( 
	\frac{VQ_s}{\lambda_{th}^3} \right)^{m_s}\nonumber \\
	=&\prod\limits_{s}\exp(c_0 V 
	Q_s)=\exp(c_0 V \mathcal{Z}_{cl}).
	\label{eqn:Zbox}
\end{align}

The box partition function turns out to be directly related to the cluster 
partition function. The box partition function is more similar to the textbook 
definition, but the textbooks typically don't include different types of 
``super-particles'' in the sum. Since in the box partition function the 
clusters of different types do not interact with each other, they are treated 
as independent and coexistent ideal gases of ``super-particles'', or clusters. 
The statistical weight of each cluster plays the role of fugacity of the 
corresponding ``super-particle''. In order to find the relationship between 
this effective fugacity and the cluster concentration, we can perform the usual 
derivative:
\begin{equation}
	c_s=\frac{1}{V}\pdv{\ln \mathcal{Z}_{box}}{\ln Q_s}=c_0 Q_s.
\end{equation}

The reference concentration $c_0$ is the same for all cluster types, therefore 
the statistical weight $Q_s$ of every cluster is directly related to the number 
concentration of the corresponding clusters. A special case of a cluster type 
$s$ is a free monomer: a cluster $i$ which consists of only one particle of 
type $i$ and no bonds so that $E_i=0$ and $S_i=0$. The statistical weight of 
such a cluster is then $Q_i=e^{\beta \mu^i}\equiv z_i$, known as fugacity. The 
concentration of such free monomers is then:
\begin{equation}
	c_{i,free}=c_0 z_i=c_0 e^{\beta \mu^i}.
	\label{eqn:c_free}
\end{equation}

Note that the concentration of \emph{free monomers} can be substantially 
different from the \emph{total concentration} of all monomers, free and bound. 
We discuss this distinction below. The relationship \eqref{eqn:c_free} can be 
interpreted as a simple nonlinear unit conversion between the chemical 
potential and the free monomer concentration. It is agnostic of what else is 
happening in the system and is thus not an equation of state.

\subsection{Yield metrics}
We show in the main text that the cluster partition function \eqref{eqn:Zcl} 
inevitably diverges at some values of chemical potential $\mu$. The apparent 
unphysicality of the partition function divergence is resolved by 
carefully considering the experimentally controllable variables. If we treat 
the vector of chemical potentials $\vec{\mu}$, or equivalently, the vector of 
free monomer densities 
$\undervec{c}_{free}$ as a free parameter, we encounter that for certain 
values the partition function diverges. This means that given the building 
block properties and interactions, there don't exist any \emph{equilibrium} 
configurations of the system with that many free monomers. We can certainly 
prepare a system configuration with an arbitrarily high concentration of free 
monomers (for example the moment when the building blocks are just placed into 
the box), but that configuration will be far from equilibrium. As it 
equilibrates, many monomers will bind up into various structures and become not 
free anymore.

The quantity conserved throughout the equilibration process is the total 
monomer concentration of each type $\undervec{c}_t$, since an 
experimentalist can put an arbitrary proportion of monomers into the box. In 
order to compute the total concentration, we need to add up the numbers of 
monomers bound up in each type of cluster. A cluster type $s$ has the number 
$\undervec{n}_s$ of each monomer type, and thus contributes $c_0 Q_s 
\undervec{n}_s$ to the total particle concentrations. The number of particles 
can be pulled out of the exponential in $Q_s$ by taking a derivative with 
respect to $\vec{\mu}$ (remember that a derivative of a scalar with respect to 
a column vector is a row vector). Fortunately, the partition function sum has 
already been performed, so the total concentration is found by a 
straightforward derivative:
\begin{equation}
	\undervec{c}_t=c_0\frac{1}{\beta}\pdv{\mathcal{Z}_{cl}}{\vec{\mu}}.
	\label{eqn:c_total}
\end{equation}

The expression \ref{eqn:c_total} is the equation of state for the 
self-assembling system, since it relates two conjugate thermodynamic variables. 
If the total concentrations $\undervec{c}^\textsf{total}$ are known, the 
chemical potentials $\vec{\mu}$ can be found. The two vectors have the same 
dimension $N$, so the expression \eqref{eqn:c_total} is a map $\mathbb{R}^N\to 
\mathbb{R}^N$, or a system of $N$ coupled equations. Since the particles of 
different types can interact with each other (that is the whole point of 
heterogeneous self-assembly), the equations are not separable and need to be 
solved jointly.

The partition function $\mathcal{Z}_{cl}$ diverges as $\vec{\mu}$ approaches a 
critical surface in $\mathbb{R}^N$. The derivative of the partition function 
then diverges even faster. This fast divergence ensures that any arbitrarily 
high values of total concentration are reached close to the critical 
$\vec{\mu}$. Conversely, if in experiment we control 
$\undervec{c}_t$ and can make it arbitrarily high, then we can 
approach the critical $\vec{\mu}$ surface arbitrarily closely but never cross 
it. If the crossing never happens and we always stay in the region where the 
partition function is analytic, there will be no observable equilibrium phase 
transition as the total concentrations are varied.

The divergence occurs because of the growth of the magnitude of terms with 
large $\undervec{n}_s$. These terms correspond to large, complex structures of 
many monomers -- but those are precisely the structures we want to assemble! 
Therefore, in order to observe high yields of the complex structures, we need 
to be in the near-divergent regime of the parameter space, and understanding 
the nature of the divergence for a given self-assembly model is incredibly 
important.

How can we then evaluate the yield of a particular structure? The simplest 
metric is the \emph{relative yield}, introduced in the 
Ref.~\cite{murugan2015undesired}. We can either designate a specific structure 
$s$ as the target, or aggregate the weights of several structures, for instance 
all that share the same topology, into a larger target. The relative yield is 
then given by:
\begin{equation}
	Y^{rel}_{targ}=\frac{Q_{targ}}{\mathcal{Z}_{cl}},
\end{equation}
where $Q_{targ}$ is either a specific $Q_s$ or sum of all $Q_s$ designated as 
the target. Physically, the relative yield answers the question: given a 
randomly-picked cluster, what is the probability that this cluster is of target 
type? Notably, the picking of clusters is done without taking their size into 
account, thus equally comparing free monomers with most complex structures.

Whether or not the relative yield is an appropriate metric depends on the 
design goals. If we interpret self-assembly as a manufacturing technique, we 
can evaluate it by the amount of target structures it successfully produces, or 
by the conversion of raw monomers into the target structure. We can compute 
such metrics by combining the notions of total concentration and the relative 
yield. We define the \emph{absolute yield} as the concentration of monomers 
bound up in the target structure:
\begin{equation}
	\undervec{Y}^{abs}_{targ}=\undervec{c}_{targ}=\frac{c_0}{\beta} 
	\pdv{Q_{targ}}{\vec{\mu}}.
\end{equation}

Note that the absolute yield tracks each type of particle separately in a row 
vector form. Similarly, we define the \emph{conversion ratio} as the percentage 
of all monomers that are bound into the target structure:
\begin{equation}
	\undervec{R}_{targ}=\undervec{Y}^{abs}_{targ} / 
	\undervec{c}_t = \left( \pdv{Q_{targ}}{\vec{\mu}} \right) / 
	\left( \pdv{\mathcal{Z}_{cl}}{\vec{\mu}} \right),
\end{equation}
where the division of vectors is carried out elementwise.

We still need to relate these yield metrics to the experimentally controlled 
parameters, i.e. the total concentrations $\undervec{c}_t$. 
However, solving the equations that relate $\undervec{c}_t$ to 
$\vec{\mu}$ is in general hard and becomes even harder and more numerically 
unstable close to a divergence point, so it is an undesirable strategy. 
Instead, we will treat the chemical potential $\vec{\mu}$ as an implicit 
parameter and plot parametric yield curves. It is important to determine the 
exact point of divergence as precisely as possible and have a fine, possibly 
non-uniform grid of 
$\vec{\mu}$ close to it since small variations in the chemical potentials there 
correspond to great changes of the concentrations. As well, to avoid sampling 
high-dimensional spaces, we can consider the situations where all of the 
chemical potentials are the same $\mu^i=\mu=const$, or follow a simple pattern 
parameterized with a single number $\mu^i=\mu+\Delta \mu^i$ for $\Delta \mu^i$ 
a constant vector.

In practice studying and visualizing the whole pattern of total concentrations 
and conversion ratios is inpractical, so instead we add up the numbers of 
particles of all types to get scalar numbers. In the main text we use the 
following scalar definitions:
\begin{align}
	c_t=&\sum\limits_{i}c_{i,t}\\
	c_{targ}=&\sum\limits_{i}c_{i,targ}\\
	R=&c_{targ}/c_t
\end{align}

\subsection{Polymer models}
Apart from binding energies and chemical potentials, the statistical weights of 
different clusters depend on their entropy. Entropy is a shorthand for the 
number of microstates that are identified as the same cluster. While we can 
tell apart two clusters based on the sequence of their monomers and the 
presence of specific bonds, we choose to treat two clusters with slightly 
different bond angles to be identical. However, the number of states that 
results from such bending depends on the size of the structure and its topology.

All structures considered in this paper are one-dimensional sequences of 
monomers: the key site of one is bound to the lock site of the next. The last 
monomer in the sequence may or may not be bound to the first one, thus the 
whole structure looks either like a linear chain, or a closed ring. Here we 
discuss 
the entropy of this chain or ring, while the next section is devoted to 
accounting for the sequence-dependent energy and chemical potential. To account 
for the bending entropy, we adopt the worm-like chain model 
\cite{doiedwards1988} described by the following Hamiltonian:
\begin{equation}
	\mathcal{H}=-P\sum\limits_{k=1}^{n-1} \hat{t}_k\cdot 
	\hat{t}_{k+1}=-P\sum\limits_{k=1}^{n-1}\cos(\theta_k),
\end{equation}
where $P$ is the bending rigidity, $n$ is the length of the chain, $\hat{t}_k$ 
is the unit vector in the direction of each monomer, and $\theta_k$ is the 
angle between two subsequent monomers. The Hamiltonian is minimized when each 
monomer has the same direction as the previous one, and deviations from this 
minimum are penalized more for larger $P$. The wormlike chain model is 
equivalent to the Heisenberg model of ferromagnetism where the unit vectors are 
mapped to spins. For a one-dimensional chain with open boundary conditions each 
of the angles $\theta_k$ can be treated as an independent integration variable, 
thus finding the partition function is straightforward:
\begin{align}
	\mathcal{Z}_{wl\;chain}^{(n)}=&\int \left( 
	\prod\limits_{k=1}^{n-1}d\Omega_k \right)e^{-\beta \mathcal{H}}\nonumber\\
	=&\left(4\pi 
	\frac{\sinh(\beta P)}{\beta P}\right)^{n-1}=e^{S_b\cdot(n-1)},
\end{align}
where $d\Omega_k$ is integration over a solid angle and $S_b$ is the bending 
entropy of each of the $n-1$ bonds. From the partition function we can easily 
extract the correlation in direction of the two adjacent monomer directions:
\begin{align}
	\expval{\cos(\theta)}_2=&-\frac{1}{\beta}\pdv{\ln 
		\mathcal{Z}_{wl}^{(2)}}{P}=\left[ \coth(\beta P)-\frac{1}{\beta P} 
	\right]\equiv c(\beta P)\\
	\expval{\cos(\theta)}_n=&c^{n-1}=e^{-(n-1)/\xi},
\end{align}
where $\xi\equiv-1/\ln c$ is the \emph{persistence length} of the chain, 
measured in 
monomers. Note that $c$ is an analytic function of bending rigidity as the 
singular parts of the two terms exactly cancel out. The large-scale statistics 
of the polymer chain with finite bending rigidity are those of a 
\emph{persistent random walk}. A persistent walk of $n$ steps of length $a$ is 
statistically similar to a independent random walk of $n'=n/\xi$ steps of 
length $a'=a\xi$. Specifically, the probability distribution for the 
displacement between the start and the end of the random walk is approximately 
Gaussian:\cite{doiedwards1988}
\begin{equation}
	p(\vec{r})=\frac{1}{(2\pi n' a'^2)^{3/2}}\exp(-\frac{3\vec{r}^2}{2 n' 
		a'^2}).
\end{equation}

A loop is a random walk on $n$ steps that ends approximately in the same place 
where it starts (so $\vec{r}=0$), up to the size of a monomer (a volume of 
$a^3$). Note that a ring of $n$ monomers has $n$ bonds as opposed to $n-1$ in 
an open chain. The partition function of the ring is then:
\begin{equation}
	\mathcal{Z}_{wl\;ring}\approx e^{S_b n} a^3 p(\vec{r}=0)=
	\frac{e^{S_b n}}{(2\pi \xi n)^{3/2}}.
	\label{eqn:entropy_loop}
\end{equation}

Compared to the linear chains, the loop partition function has an additional 
power law decay factor with exponent $-3/2$. The decay is driven by the fact 
that it gets harder and harder for a long random walk to encounter its own 
beginning. For this reason, long loops are always entropically suppressed 
compared to long chains. However, loops have one more bond that can be 
energetically favorable. We will see that this entropy-energy tradeoff is one 
of the main design limitations of the self-assembly system.

Note that the expression \eqref{eqn:entropy_loop} only holds for loops long 
enough to be able to bend on themselves, that is $n>\xi$. For very short loops, 
the bending energy will be prohibitive for the loop to form at all. The 
simplest way to deal with such short loops is to manually exclude them from the 
computation of the full self-assembly partition function, as we will do in the 
next section.

One way to reduce the entropy penalty is to use building blocks with binding 
sites that are not strictly at the opposite sides of the block, but at a finite 
angle. The lowest bending energy configuration will then correspond to a bond 
bend by that specific angle, thus directly building curvature into the 
assembling structure and favoring rings over chains. However, we will not 
consider this case in the following calculation in order to highlight other, 
more fundamental design trade-offs.

%\bibliography{biblio}
%apsrev4-2.bst 2019-01-14 (MD) hand-edited version of apsrev4-1.bst
%Control: key (0)
%Control: author (8) initials jnrlst
%Control: editor formatted (1) identically to author
%Control: production of article title (0) allowed
%Control: page (0) single
%Control: year (1) truncated
%Control: production of eprint (0) enabled
%

\end{document}